\documentclass[fleqn, pra, twocolumn, letterpaper, groupaddresses, superscriptaddress]{revtex4}

\usepackage[utf8]{inputenc}
\usepackage[T1]{fontenc}
\usepackage{times}
\usepackage{subfigure}
\usepackage{amsmath,amsfonts,amssymb,amsthm}
\usepackage{algorithm,algpseudocode}
\usepackage{mathtools}
\usepackage{commath} 
\usepackage{comment}
\usepackage{xcolor}
\usepackage{placeins}
\usepackage{booktabs}
\usepackage{hyperref}

\usepackage{enumitem} 
\newlist{condenum}{enumerate}{1} 
\setlist[condenum]{label=\bfseries Condition \arabic*., 
	ref=\arabic*, wide}

\newcommand{\vect}[1]{\boldsymbol{\mathbf{#1}}}

\begin{document}

\title{Quantum Molecular Unfolding}

\author{Kevin Mato }
\affiliation{Polytechnic of Milan, Milan, Italy}
\author{Riccardo Mengoni  \footnote{email: r.mengoni@cineca.it}}
\affiliation{CINECA,  Bologna,   Italy}
\author{Daniele Ottaviani}
\affiliation{CINECA,  Bologna,   Italy}
\author{Gianluca Palermo}
\affiliation{Polytechnic of Milan, Milan, Italy}

\begin{abstract}

Molecular Docking (MD) is an important step of  the  drug discovery process which aims at calculating the preferred position and shape of one molecule to a second when they are bound to each other.
During such analysis, 3D representations of  molecules are manipulated according to their degree of freedoms: rigid roto-translation and fragment rotations along the rotatable bonds.
%
%
In our work, we focused on one specific phase of the molecular docking procedure i.e. \emph{ Molecular Unfolding (MU)}, which is used to remove the initial bias of a molecule by expanding it  to an unfolded shape. 
The objective of the MU problem is to find the  configuration that maximizes the molecular area, or equivalently, that maximizes the internal distances between  atoms inside the molecule. 
We propose a quantum annealing approach to MU by formulating it as a High-order Unconstrained Binary Optimization (HUBO) which was possible to solve on the latest D-Wave annealing hardware ({2000Q and  Advantage}).
Results and performances obtained with quantum annealers are compared with  state of art classical solvers.
\\

\end{abstract}

\maketitle

\section*{Introduction}

Drug Discovery~\cite{Lyne2002} is a process that includes several phases, from virtual \emph{in silico} simulations to \emph{in-vitro} and \emph{in-vivo} experimentation. 
Molecular Docking~\cite{Meng2011} (MD) is an important step of  the  drug discovery process which aims at calculating the preferred position and shape of one molecule to a second when they are bound to each other.
The computational resources needed to address MD related tasks are usually quite large, as problems of this kind  involve several degrees of freedom and rapidly growing dimensionality.

In recent years, the field of Quantum Computing~\cite{nielsen_chuang_2010} (QC) has undergone significant developments  from  both hardware and algorithms points of view.
In particular, we are living in the so-called NISQ~\cite{Preskill2018quantumcomputingin} era in which it is not clear whether the quantum devices currently available are actually capable of producing better or comparable results compared to classical methods.

In this work, we explored the possibility of  using a QC technique called Quantum Annealing~\cite{7055969,babej2018coarsegrained,Marchand2019} (QA)  in the virtual screening phase of MD.
In particular, we focused on the Molecular Unfolding (MU) process, which is the first step in geometric molecular docking techniques.
MU aims at finding the molecular configuration that maximizes its volume, or equivalently, that maximizes the internal distances between atoms inside the molecule. 

Our aim was to develop a  Quantum Molecular Unfolding approach and execute it on the   latest quantum annealing hardware (D-Wave Advantage and 2000Q) in order to  understand the capabilities of these devices and compare their performances to  state-of-art molecular docking methods.

The remainder of the paper is structured as follows.
In the first section, the topic of molecular docking is introduced, followed by a section where the problem of Molecular Unfolding is discussed. In  Sec.3, concepts of quantum annealing are illustrated while in  Sec.4  a binary optimization formulation of the problem is given.  Later in the fifth section, the dataset and   pre-processing steps employed before the unfolding phase are presented. This is followed in Sec.6 by  a review  of  state of the art classical algorithms employed to compare solutions obtained with  the quantum devices. Finally, in the last two sections, results regrading the embedding  on the QPUs are presented  as well as a detailed performance comparison between the different methods.

\section{Preliminaries on Molecular Docking}

In-silico approach to drug discovery~\cite{Lyne2002} can be viewed as a multi-tiered process  that encompasses several sequential computational techniques  with the aim of screening virtual libraries of the order of billions of compounds for the most suitable molecules to forward to later experiments.  

Molecular Docking~\cite{Meng2011} (MD) is an essential step in virtual screening which is used  to simulate the atomic interactions of  a ligand  inside a protein binding site, in order to highlight their possible biochemical reactions and predict whether a stable complex could be formed.
\begin{figure}[htbp] 
	\centering
	\includegraphics[width=0.45\textwidth]{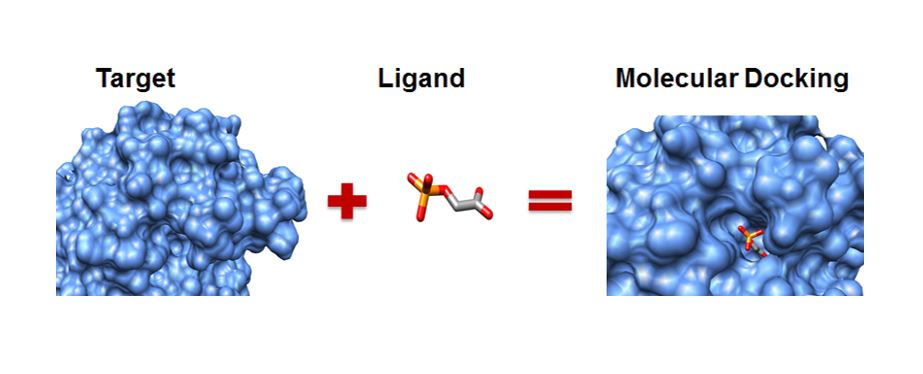}
	\caption{Visual representation of Molecular Docking  for a ligand inside a
		protein pocket, similar to the relation between locks and keys.}
	\label{fig:background:Lock and Key}
\end{figure}
\\
\\

The MD process is divided in two main tasks: 
\begin{itemize}
	\item Detection of three-dimensional \emph{poses} of the ligand, i.e. valid ligand conformations, positions and orientations inside the active site  of the protein (usually  called \emph{pocket}).

	\item Ranking of the poses  via a scoring function.  Usually the lower  the docking score, the better  the resulting  \emph{binding affinity}.
\end{itemize}

Since  electrons inside atoms are repulsive to each other, their interaction  affects both molecular shape and  reactivity. Therefore,  by controlling the molecular shape, it is possible to  predict the ligand  reactivity to a protein's pocket as well as the energy cost of such configuration. This direct connection between molecular conformation and binding affinity is what enables geometrical scoring functions~\cite{Gadioli2020}.

In our approach, the docking process considers the pocket as a rigid structure, while the ligand is a flexible set of atoms. Furthermore, from a strictly geometrical interpretation, the ligand can  be seen as a set of chemical bonds (or edges) with a fixed length, where only a  subset of edges are  \textit{rotatable}.

\begin{figure}[htbp] 
	\centering
	\includegraphics[width=0.4\textwidth]{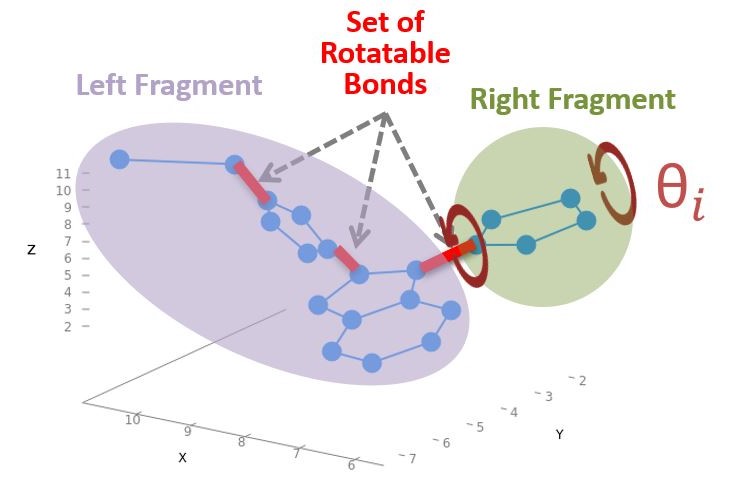} 
	\caption{Fragments and rotatable bonds}
	\label{fig:background:Fragments and rotatable bonds.}
\end{figure}

Rotatable bonds (or torsionals) are bonds that split the molecule in two nonempty disjointed fragments, when virtually removed.  Consequently, such fragments  can rotate independently from each other around the axis of the rotatable bond. 
These  concepts are graphically reported in Fig.\ref{fig:background:Fragments and rotatable bonds.}, where the rightmost rotatable bond is splitting the molecule in left and right fragment.
Finally,  note that the number of fragments in a molecule is usually double the number of  rotatables.

\FloatBarrier

In general, it is possible to point out three main phases in MD: \emph{Ligand expansion}, \emph{Intial Placement} and\emph{ Shape Refinement} inside the pocket.

Here we focus on  the ligand expansion phase, which is essential for improving  docking. In fact,  an initial pose of the ligand that is set a priori  may introduce shape bias  affecting  the final quality of the docking.  Molecular Unfolding (MU) is the technique used for removing such initial bias  by expanding the ligand to an unfolded shape.

\section{Molecular Unfolding Problem Definition}
\begin{figure}[htbp]
	\centering
	\includegraphics[width=0.4\textwidth]{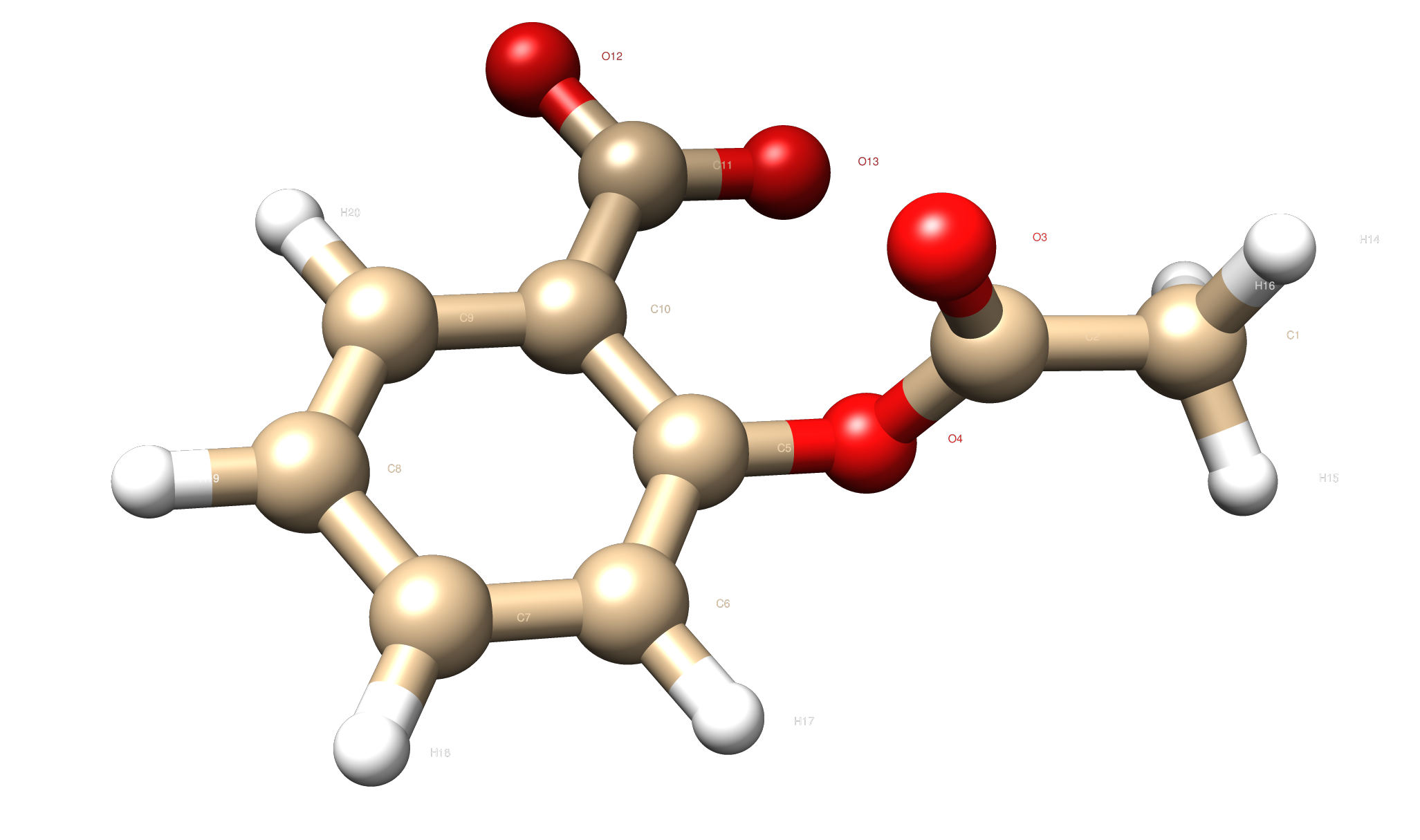}\ \ \
	\includegraphics[width=0.45\textwidth]{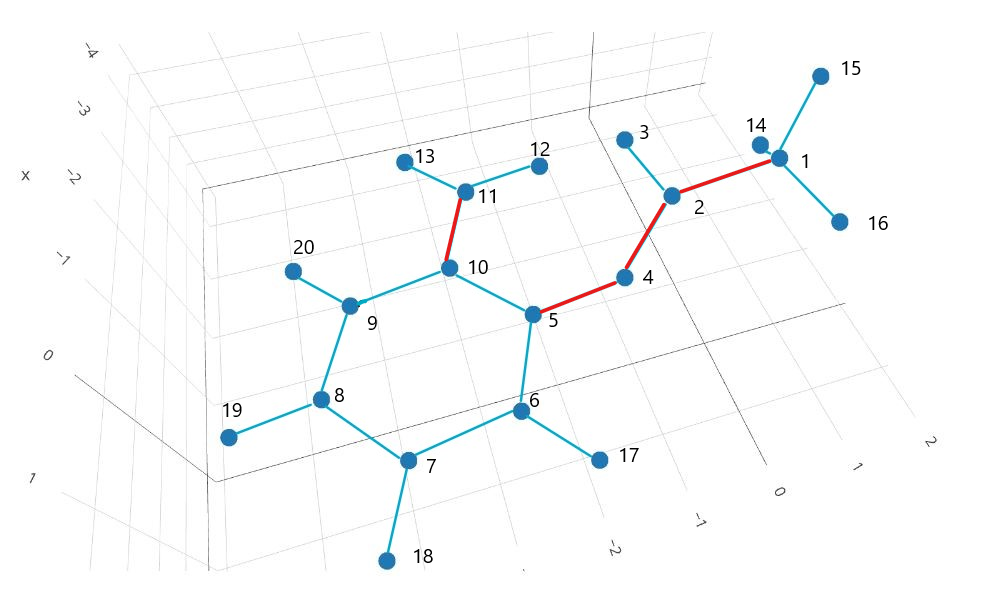}
	\caption{ Illustration of a simple organic molecule. The molecular structure presents 4 rotatable bonds, highlighted  in red in the graph representation.}
	\label{fig:test_molecule2}
\end{figure}

The objective of the MU problem is to find the unfolded shape of the ligand or,  in other terms, the torsion configuration that maximises the molecular volume expressed as   total sum of  internal distances between pairs of atoms in the  ligand. The starting point is the folded molecule defined by a  set of atoms together with its  fixed and rotatable bonds. 
We can assign to each  rotatable bonds $t_{i}$ a variable corresponding to the angle $\theta_{i}$  which is responsible for fragment rotation.
As a convention, we  identify the ordered set of rotations by a vector:
\begin{equation}\label{eq:t_vec}
t=\begin{bmatrix} \theta _{1}, \theta _{2},  \ldots  , \theta _{n} \end{bmatrix}
\end{equation}
where each torsion  $\theta_i$ around the bond's axis can assume  values $\left[ 0, 2\pi \right)$.
Given a molecule, it is possible to construct the associated graph as shown in Figure~\ref{fig:test_molecule2} where  edges represent bonds (which  can not contract).  Torsional bonds are depicted in red in the image below.

\FloatBarrier

The objective for the MU problem can be expressed in mathematical terms as follows: given a molecule,  find the  torsional configuration 
\begin{equation}
\centering
\vect{t^{unfold}} = [\theta_1^{unfold},\, \dots , \theta_M^{unfold}]
\end{equation}

such that the following quantity is maximised

\begin{equation} \label{eq:D}
\centering
D({\bf \vect{t}}) = \sum_{\substack{a,b \in \mathcal{M} \\ a \neq b}}  D_{ab}({\Theta})^2\,
\end{equation}

The function $ D_{ab}( \Theta) $ denotes the distance between two different atoms $ a $ and  $ b $ belonging to the molecule $ \mathcal{M} $ while $ D( \bf{ \vect{t} }) $ is  the sum of all square  distances between  atoms.

The rationale behind this choice is that the objective function is simple and relies just on the geometry of the molecule.  Each distance  $ D_{ab}( \Theta) $  only depends  on the angles induced by those rotatable bonds that appear in the shortest path connecting atom $a$ to $ b$ in the graph.

It is worth noticing that  it is not necessary to calculate all the pairwise distances that are expressed in equation \ref{eq:D}. In fact, it is possible to apply  simplifications to  $ D_{ab}(\Theta) $, as expressed by the following two conditions:

\begin{condenum}
	\item The shortest path between two atoms must contain at least one torsional. This is due to the fact that if two atoms belong to the same rigid fragment,  their relative position  never changes. \label{cond: conditions4Distance1}
	\item The shortest path connecting two atoms must have at least three edges. This is a consequence of the fact that atoms connected by only one or two  bonds  have always the same distance, even when  bonds are rotated. \label{cond: conditions4Distance2}
\end{condenum}

A useful mathematical representation for rotations are rotation matrices, $R(\theta_i)$, which are function  of the  angle $ \theta_i $  associated to torsion $ T_i $ .  
If two atoms are connected via multiple torsional, the rotation matrix  $R({\Theta})= R(\theta_i,\theta_j, \dots ,\theta_k)  $,
characterizing their relationship is   
an ordered multiplication of single torsion rotation matrices
\begin{equation} \label{eq:R}
R({\Theta})=R(\theta_i,\theta_j, \dots, \theta_k) = R(\theta_i)\times R(\theta_j) \times ... \times R(\theta_k)
\end{equation}
Each rotation  $R(\theta_i)$ is expressed via a $4\times4$ rotation matrix  acting on atomic coordinates.  The extremes of a bond connecting  two atoms are identified by the coordinates of the atoms themselves.

\begin{figure}[htbp] 
	\centering
	\includegraphics[width=0.30\textwidth]{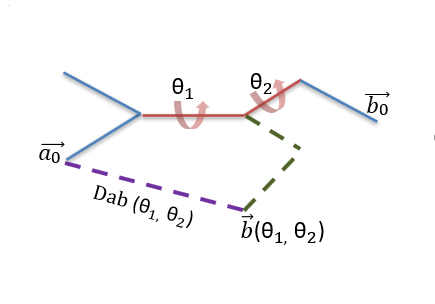}
	\caption{ Visual representation of the action of two rotations $\theta_1$ and $\theta_2$ on the distance expressed in eq.
		\ref{eq:Euclidean_dist}.}
	\label{fig:distance_according_to_rotation}
\end{figure}

Consider now the distance $ D_{ab}( {\Theta}) $ and 
the initial positions of atoms  $ a $ and $ b  $   which are  identified with  $\vec{a_0}$ 
and
$\vec{b_0}$ respectively. 
Relative positions are obtained by fixing one of the two atoms while  rotating the other as follows
\begin{equation}
\vec{a}= \vec{a_0} \ \, \ \ \ \ \  \  \vec{b}=R({\Theta}) \vec{b_0}
\end{equation}
Hence, we can rewrite the single contributes  in equation \ref{eq:D} using the Euclidean distance as
\begin{equation}\label{eq:Euclidean_dist}
D_{ab}( {\Theta})^2 = \| \vec{a_0} - R({\Theta}) \vec{b_0}\|^2
\end{equation}
Figure~\ref{fig:distance_according_to_rotation} shows how the    relative positions change as function of the angles assumed by the two rotatable bonds.

\FloatBarrier

\section{Quantum Annealing}
Quantum annealing (QA) is a meta-heuristic technique for addressing challenging combinatorial problems.
It can be seen as a  variation of the \emph{Simulated Annealing (SA)}~\cite{Kirkpatrick1983} algorithm where quantum fluctuation, instead of thermal effects, are employed in the exploration of the configuration space.
Quantum annealing devices, like those   implemented  with \emph{superconductive qubits} by the Canadian  company \textit{D-Wave}, are able to  solve  combinatorial  problems expressed in   Quadratic Unconstrained Binary Optimization (QUBO) terms. 
A generic QUBO problem  is defined as
\begin{equation}\label{eq: Ising_problem}
O(x)=\sum_{i}h_i {x}_i+ \sum_{i>j}J_{i,j} x_i x_j
\end{equation}
where $ x_i \in \left\lbrace 0,1 \right\rbrace  $  are binary variables while  $ h_i $ and  $ J_{ij} $ are parameters whose values encode the optimization task  in such a way that the  minimum of $ O(x) $ represents the solution to the optimization problem.

In general,  a given  optimization problem can be easily translated  in binary combinatorial terms as a a high-order quadratic unconstrained binary optimization (HUBO) problem. 
\begin{equation}\label{eq: Ising_problem}
O_{Hubo}(x)=\sum_{i}\alpha_i {x}_i+ \sum_{i,j}\beta_{i,j} x_i x_j+  \sum_{i,j,k}\gamma_{i,j,k} x_i x_j x_k +...
\end{equation}

Fortunately, it is always possible to convert  HUBOs into QUBOs.  The trick used is to add new ancillary binary variables and  substitute  high-order terms with  sums of quadratic expressions (made up with   original binary variables and  new ones) in order to preserve  local and global minima~\cite{mengoni2020breaking}.

As an example, a cubic term like $ \pm x_1\cdot x_2\cdot x_3 $ can be divided using the ancillary binary variable $a_1$ as
\begin{equation}\label{eq: make_quadratic_rel}
\pm x_1 \cdot x_2 \cdot x_3 \rightarrow  \pm a_1 x_3 + 2 (x_1 x_2 - 2 x_1 a_1 - 2 x_2 a_1 + 3 a_1 )
\end{equation}
I practice, one can decide to build a custom function for quadratisation or exploit the functions present in D-Wave software offer such as \textit{make\_quadratic}.

\section{HUBO Formulation}
As a first step, it is necessary to formulate  the MU problem in binary optimization terms in order to employ a quantum annealing approach.

Let us consider a discretisation of the angle $ \theta_i $ associated to a rotatable bond $T_i$ into $d$ possible values
\begin{equation}
\theta_i= [\theta_i^1 , \theta_i^2 ,  \theta_i^3,..., \theta_i^d ]
\end{equation}
As a consequence,  continuous functions of the angles like sine and cosine can be  discretised as well in $d$  values
\begin{align}
\sin(\theta_i)& = [\sin(\theta_i^1) , \sin(\theta_i^2) ,  \sin(\theta_i^3) ,..., \sin(\theta_i^d)   ]\\
\cos(\theta_i)& = [\cos(\theta_i^1) , \cos(\theta_i^2) ,  \cos(\theta_i^3) ,..., \cos(\theta_i^d)   ]
\end{align}
Moreover, since torsional angles can assume  only one value at a time, it is necessary to   associate each $\theta_i$ with only one value among all the possible $\theta_i^k$. A {\emph{One-Hot Encoding}} strategy is adopted for each  torsional $T_i$, to which are  assigned  binary variables $x_{ik}$,  with  $1 \leq k \leq d$,  such that
\begin{equation}\label{eq:system_classical_constraint_onehot}
x_{ik} = \begin{cases} 
1 & \text{if } \theta_i= \theta_i^k; \\
0 & \text{otherwise.} 
\end{cases}
\end{equation}
Definition \ref{eq:system_classical_constraint_onehot} tells us that only one among the binary variables can be assigned to a one or truth value. This means that  their sum  is always equal to one: 
\begin{equation}\label{eq:sum_classical_constraint_onehot}
\sum_{k=1}^d x_{ik} =1
\end{equation}
Note that equation \ref{eq:sum_classical_constraint_onehot} must hold for all  rotatables. In order to convert such algebraic constraint into QUBO, the following term has to be considered:
\begin{equation}\label{eq: squared-sum-hard} 
\sum_{i} \left( \sum_{k=1}^d x_{ik} -1 \right)^2
\end{equation}
Constraint of eq. \ref{eq: squared-sum-hard} is  usually  called a \emph{hard constraint} because it is mandatory to satisfy  this term i.e. minimize it in the combinatorial optimization.

Regarding those functions that appear in the rotation matrix $R({\theta_i}) $   such as 
$\sin(\theta_i)$  and $ \cos(\theta_i) $, they can be  expressed  as
\begin{align}\label{eq: one-hot constraint}
\sin(\theta_i)&= \sum_{k=1}^d \sin(\theta_i^k)\  x_{ik} \\ 
\cos(\theta_i)&= \sum_{k=1}^d \cos(\theta_i^k)\  x_{ik}
\end{align} 
Therefore,  a rotation matrix $R({\theta_i}) $  becomes  a  function of all the binary variables  $x_{ik}$ needed to represent the angle $\theta_i$
\begin{equation} 
R({\theta_i}) =R(x_{i1}, x_{i2}, ..., x_{id})
\end{equation}
For a generic rotation we have
\begin{align} 
\nonumber
R({\Theta})& = R(\theta_i)\times R(\theta_j) \times ... \times R(\theta_k)\\
\nonumber
&=R(x_{i1}, x_{i2}, ..., x_{id})\times R(x_{j1}, x_{j2}, ..., x_{jd})  \times ...\\
&... \times R(x_{k1}, x_{k2}, ..., x_{kd})
\end{align}

Note that, for what concerns the granularity of the rotations, in order to obtain a precision of $\Delta\theta_i=~\theta_i^{k+1}-~\theta_i^k $ in  the representation of angle $\theta_i$,  the  number of variables $x_{ik}$ needed for each torsion is given by
\begin{equation} \label{eq: discretization}
d =\dfrac{2 \pi}{\Delta\theta_i}= \dfrac{2 \pi}{\theta_i^{k+1}-\theta_i^k}
\end{equation}
Given a number $ M $  of torsional bonds, the total number of binary variables is 
\begin{equation} 
n=d\times M=\dfrac{2 \pi}{\Delta\theta_i}\times M
\end{equation}
The general form of the HUBO optimization function can be written as
\begin{equation}\label{eq: HUBO}
O(x_{ik})=A_{const}\sum_{i} \left( \sum_{k=1}^d x_{ik} -1 \right)^2 -  \sum_{\substack{a,b}} D_{ab}( {\Theta})^2
\end{equation}
In the last equation, the second term identifies the \emph{optimization term}, which has a minus sign in front because it should be maximized (while the whole expression is minimized). The first term instead represent the \emph{hard constraint}, where the parameter $ A_{const} $ is a penalty scalar that  modulates its strength.
Finally, it is worth noticing that, given a molecule with $M$ torsionals, the highest order term appearing in eq.\ref{eq: HUBO} is $2M$, where the number 2 comes from the  squared distances.

\section{Dataset and Pre-processing}
\subsection{Ligand Dataset}
The original ligand dataset in our possession is made up of 118 molecules, with a number of atoms ranging form 20 to more that 120   and a number of torsionals going from few units to about 50.

As shown in Figure \ref{fig:proof_expansion} below,  it is possible to see that 60\% of the ligands in our database  present  10-20 fragments, which means 5-10 rotatable bonds. The number of atoms is peaked about the   40-50  interval.
For this reason, we decided to concentrate our research only on molecules that had at most  $12$ rotatable bonds and an increasing number of atoms going from $20$ to $50$.


Since our study aimed at highlighting the change in problem complexity  by increasing the number of torsionals, we considered only a subset of torsionals at a time, out of 12 possible rotatable bonds.

\begin{figure}[htbp] 
	\centering
	\includegraphics[width=0.45\textwidth]{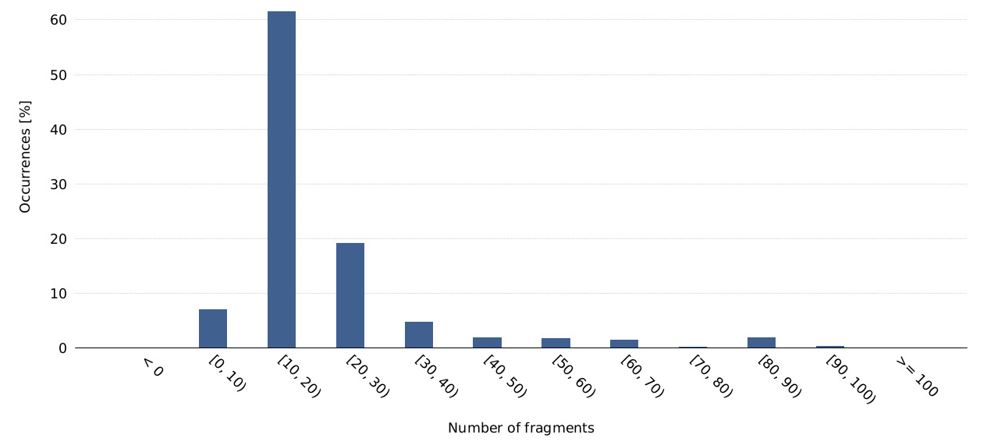}
	\includegraphics[width=0.45\textwidth]{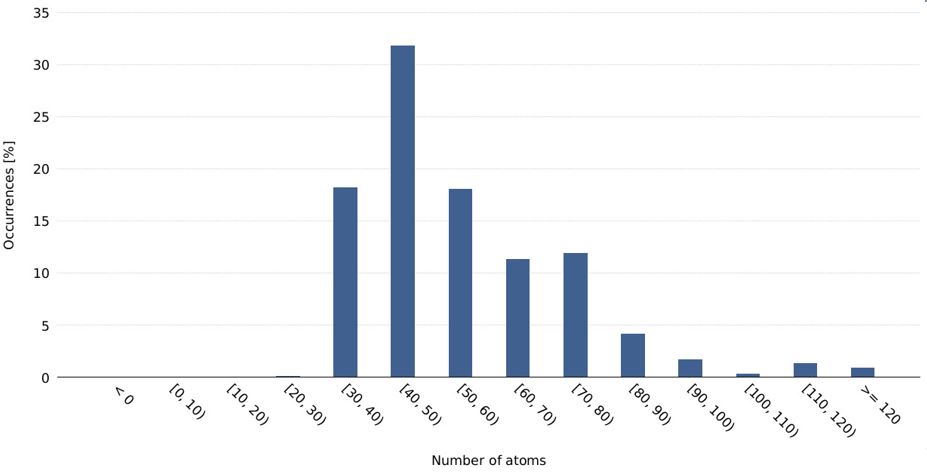}
	
	\caption{Analysis of the distribution for number of atoms and number of fragments.}
	\label{fig:proof_expansion}
\end{figure}
\FloatBarrier

\subsection{Pre-processing Phase}
The pre-processing   starts with the parsing of the information about the  molecule via analysis of its .MOL2 file.
A graph containing a three-dimensional description of the ligand together with bond type (bonds could be single, double, triple, amide or aromatic) is constructed.

Two  simplifications related to chemical-physical properties of ligands  are applied in the  pre-processing phase: one is due to the fact that only single and aromatic bonds can be  rotatable, while the other concerns the removal of terminal hydrogen atoms whose distances with the rest do not affect the total sum of internal distances. 
Note that the removal of  terminal hydrogens may trigger the generation of conformations where the final position of  atoms do not satisfy their minimum Van Der Waals distance. Such invalid shapes are discarded in post-processing. 

\begin{figure}[htbp] 
	\centering
	\includegraphics[width=0.45\textwidth]{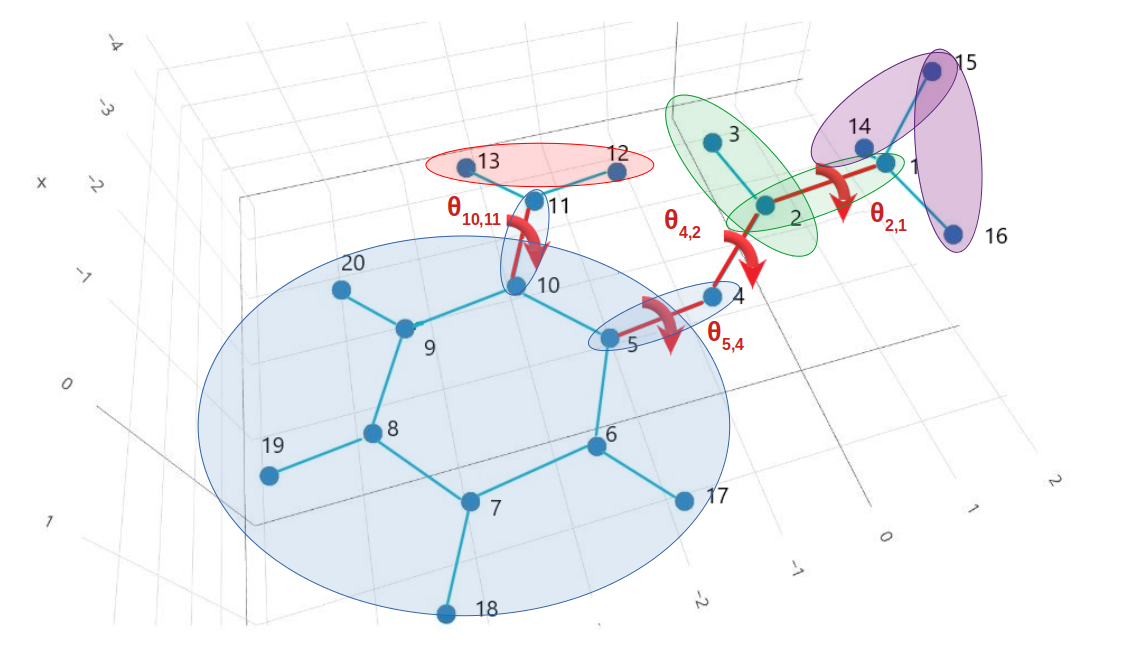}\ \ \ 
	\caption{Atoms belonging to the same fragment, depicted using colors, are affected by the same  rotatables influence set.}
	\label{fig:Rotables_influence_set_example}
\end{figure}

After this,   the  graph associated to the molucle is divided into fragments as shown in Figure \ref{fig:Rotables_influence_set_example}.
At this stage, the betweenness centrality~\cite{Freeman1977}, defined in  equation \ref{eq: betweeness},  is employed to obtain an ordering of the atoms within the  graph. By consequence, this procedure also specifies an ordering of rotatable bonds.
\begin{equation}  \label{eq: betweeness}
\mathrm{ Betweeness \ centrality\ of\ atom\ \textit{v} : }\ \ 
\sum_{v \neq s \neq t} \frac{\sigma_{st}(v)}{\sigma_{st}} 
\end{equation}
Equation \ref{eq: betweeness} is calculated for each atom $v$ as the sum of all the possible shortest paths connecting atoms $s$ and $t$ which  also cross $v$, divided by the total number of shortest paths between $s$ and $t$.
The atom with the greatest centrality is chosen as  centre of the graph and origin of the torsional ordering.

Each fragment is identified as a set of atoms influenced by the same set of rotatable bonds,  called \emph{rotatables influence set}.
The definition of rotatables influence set is the following
\begin{equation}  \label{eq: inf_set}
\mathrm{ Rotables\ influece\ set\ :\ }\ \
I_{s} = E_{C_{a,a_{k}}} \cap E_{R}
\end{equation}
$C_{a}$ = atom\ with\ greatest\ betweeness\ centrality \\
$ E_{R}$ = Rotatable\ bonds\ \\
$E_{C_{a},a_{k}}$ = Bonds\ on\ the\ shortest\ path\ $\sigma_{C_{a},a_{k}}$ \\ \\
Atoms of the same fragment are mapped to a unique shortest path, which is composed by a set of edges belonging to the influence set. 

%

%

Finally, contributions appearing in equation \ref{eq:Euclidean_dist} are obtained for each fragment, where distances between  atoms  are calculated only once and  respecting  conditions \ref{cond: conditions4Distance1} and \ref{cond: conditions4Distance2}. Distances between atoms belonging to  the same fragment are fixed, hence  not considered.
%


%

\section{Model Development}

\subsection{Distance Simplification}

The first simplification   concerns the calculation of  contributions in equation \ref{eq:Euclidean_dist}, as the number of terms  increases quadratically with the number of atoms involved.
Hence,  instead of measuring the Euclidean distance between each atom $a$ in  fragment $A$ and  each  atom $b$ in  fragment $B$, only distances between  different fragments have been considered. In particular, each fragment was identified with the coordinates of the median atom inside the fragment.

Partial contributions for single fragments are summed  up together in the HUBO as in  equation \ref{eq: HUBO}.
Note that such optimization function  is a non-linear symbolic expression where  symbolic variables representing torsional angles  appear in the rotation matrix in the form of   trigonometric functions as defined in equation \ref{eq: one-hot constraint}.

\subsection{Coarse-grained Rotations}
The number of variables needed to express the MU problem in combinatorial optimization form depends on the discretisation of the angles, as shown in eq. \ref{eq: discretization}. It is therefore essential to understand how to chose an optimal  granularity for the torsionals.
In this regard, previous knowledge about the dataset was used  to obtain approximate  solutions achieving an acceptable quality.
\begin{figure}[h] 
	\centering
	\includegraphics[width=0.48\textwidth]{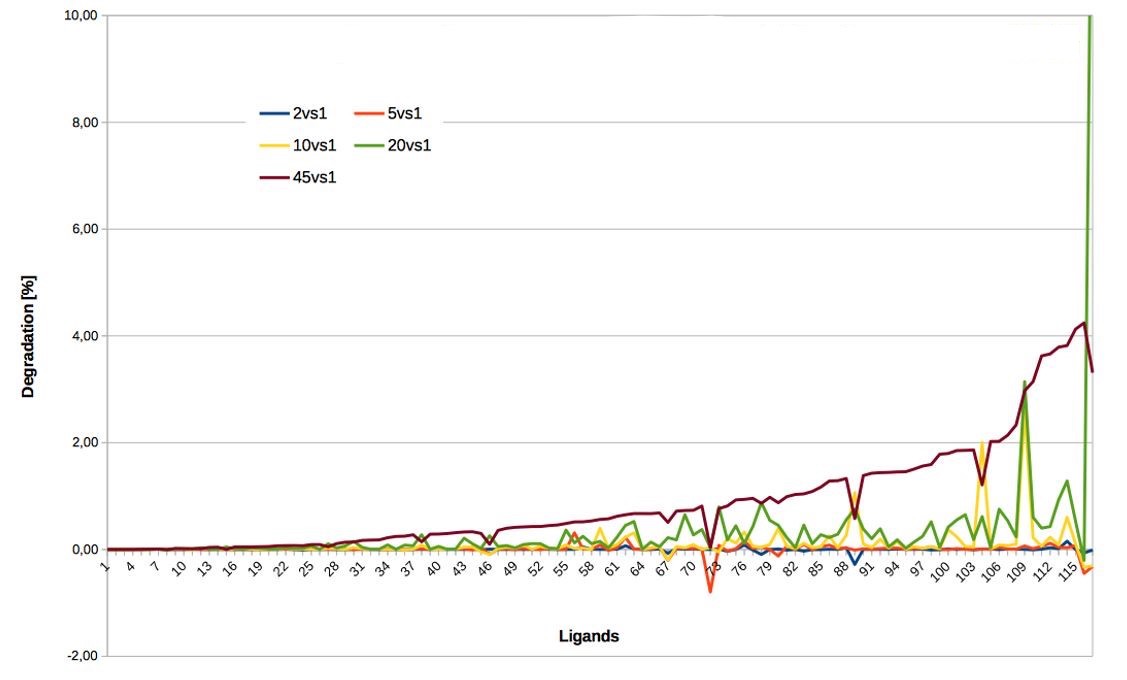}\ \ \ 
	\caption{Analysis of the  degradation by transitioning to a coarse-grained rotation. }
	\label{fig:45vs1}
\end{figure}
Our dataset  is displayed in Figure \ref{fig:45vs1}, where molecules are  ordered by increasing volume. The vertical axis shows the percentage of degradation in the molecular volume when the unfolding is made with an angle greater than 1 degree. The lines drawn in the plot identify granularities of 45, 20, 10, 5 and 2 degrees. The surprising result is that the degradation is always below 4\% for  rotations at 45 degrees with average around 1.7\%.

For this reason, even though we introduced the possibility of using different angular discretisations, as shown in eq. \ref{eq: discretization},  we focuses on  a granularity of  45 degrees. 
The combinatorial optimization  would only require  8 binary variables  for each rotatable bond to identify its angle while  keeping the final quality loss acceptably low.
\subsection{Threshold Approximation}

The generation of the HUBO phase  is one of the most computationally intensive, since we are using a symbolic framework. The challenges that arise  are strictly related to the enormous amount of memory used and the unparalleled implementation of the symbolic engine.
Hence, a  straight derivation of the HUBO becomes unrealistic when increasing problem size.

For this reason, a  last simplification was introduced, which  consists in eliminating the  optimization terms in \ref{eq: HUBO}  (only those not related to the hard constraints) whose   coefficient is strictly less than a certain threshold value.

\begin{figure}[h] 
	\centering
	\includegraphics[width=0.45\textwidth]{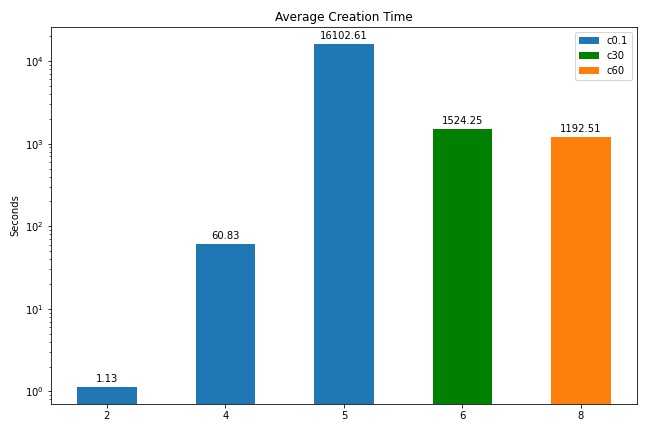}
	\caption{Average creation time for HUBOs. We were able to generate HUBOs without the threshold approximation, up to 5 torsionals. In the case of 6 and 8 rotatable bonds, the threshold was increase to the vale of 30 and 60 respectively. }
	\label{fig: HUBOtime}
\end{figure}

The effect of such threshold approximation (also called \textit{chop}) was to speed up the computation of the HUBOs, as shown in Figure \ref{fig: HUBOtime},  and decrease the number of terms in the QUBO function in order to match the quantum annealing hardware specifications and limitations such as number of qubits (i.e. binary variables) and connectivity (i.e. quadratic terms) of the problem.
Our study doesn't go beyond the  8 rotatable bonds as  HUBO construction involving 10 torsionals exceeded the time limit  with the current setup and software.



\FloatBarrier

\section{Altenative Search Strategies}
In the following, classical optimization strategies  used to compare   quantum annealing results  are presented.

\subsection{Random Search}
Random Search~\cite{random_search}  algorithm works by picking up randomly the candidate solution from a uniform distribution.
A big strength  of this method is  the high parallelization and the possibility of using  Bayesian techniques or probability distributions that are more suitable for a given problem.

\begin{algorithm}
  \caption{Random Search}
\label{Random Search}
	
	\begin{algorithmic}[1]

\Require $A \ =\ Angles,\ Poly = \ D_{ab}( {\Theta})^2 , \ T = Torsional \ Bonds \ Set$\
\Ensure  ${\gamma_{max}} : \mathbf{max}\ D_{ab}( {\gamma_{max}})^2$ \
\State $  | T | = m $ \ 
\State $ \Gamma \ =\ A_1 \times A_2 \times... \times A_m$ \ \Comment{Cartesian product of the set of angles - m times.}
\While{ $t \ < \ MaxTime$ } \\
\ \ \ $ \gamma = \ pick\_rand\_uniform(\ \Gamma \ ) \ $\\
\ \ \ $Evaluate\ Poly(\Theta = \gamma)$ \ \Comment{Evaluate assignments and record the optimum.}
\EndWhile
\State return $[\  \gamma_{max}, \ max \ Poly ]$\
		
	\end{algorithmic}
\end{algorithm}

\subsection{GeoDock Search}


The GeoDock search is a greedy algorithm which considers a single torsional at a time and makes a decision locally for each rotatable bond. The algorithm usually does not lead to an optimal solution but it can reach good approximate solutions in a reasonable time.

This algorithm is the only one that is not employing  symbolic computation. Each bond is rotated independently, starting from the most central and internal bond with respect to betweenness centrality.     Only the conformation that improves the volume is recorded as a solution. The whole process is repeated a number of times or until a plateau in the improvement is reached.

\begin{algorithm}
	\caption{GeoDock Search}
	
	\label{GeoDock-inspired Search}
	
	\begin{algorithmic}[1]
		\Require $A \ =\ Angles,\ Poly = \ D_{ab}( {\Theta})^2 , \ T = Torsional \ Bonds \ Set$\
		
		\Ensure  ${a_{max}} : \mathbf{max}\ D_{ab}( a_{max})^2$ \
		
		\State $  | T | = m $ \ 
		
		\For{ $t_i,\theta_i \ in \  T_{ordered}$ }  \Comment{Ordered by betweenness centrality.}
		\For{ $\ a \ in \ A \ $ } \\
		\ \ \ \ \ \ \ $Evaluate\ Poly(\theta_i \ = \ a )$ \ \Comment{Record intermediate optimums.}
		
		\EndFor
		\EndFor
		\State return $[\  \underline{a}_{max}, \ max \ Poly ]$\
		
	\end{algorithmic}
	
\end{algorithm}

\subsection{Simulated Annealing}

Simulated annealing (SA) is a meta-heuristic technique,  useful in the search of the optimum in large solution spaces.

\begin{algorithm}[H]
	\caption{Simulated Annealing }
	
	\label{Simulated Annealing Search}
	
	\begin{algorithmic}[1]
		\Require $A \ =\ Angles,\ Poly = \ D_{ab}( {\Theta})^2 , \ T = Torsional \ Bonds \ Set, \ N_{max} \ = \ Total\ Iterations$\
		
		\Ensure  ${a_{max}} : \mathbf{max}\ D_{ab}( a_{max})^2$ \

		\State $T\gets T_{max}$
		\State $a_{max}$ = NULL
		
		\While{$T>T_{min} \  and \  N<N_{max}$}
		\State $next$ = $pickNeighbour(T, best)$
		\State $\Delta E $ = $Poly(next)-Poly(a_{max})$
		\State $r$ = $randomNumberUniform(0,1)$
		\If{$\Delta E < 0$}
		\State $a_{max} = next$  \Comment{The neighbour is accepted as the new optimum.}
		\ElsIf{ $r< e^{ \frac{-\Delta E}{k_b * T}  } $}
		\State $a_{max}$ = $next$   \Comment{Moving to a neighbour.}
		\EndIf
		\State $T$ = $geometricDecrease(T)$ 
		\EndWhile
		
		\State return $[\  \underline{a}_{max}, \ max \ Poly ]$\
		
	\end{algorithmic}
	
\end{algorithm}
Since we can visualize the solution space as points belonging to a graph,  SA starts with an initial guess $s_c=s_i$ with its evaluation $f(s_i)$. If the neighbour has a lower evaluation, we take it as the current solution $s_c$. 
If the neighbour has a higher evaluation, we accept it with a probability written as:
\begin{equation}\label{eq:SAacceptance}
P_{accept} = \min(1,\  e^{-\frac{Poly(neigh.)-Poly(s_c)}{T}})
\end{equation}
where T is the temperature that starts from a high value and decreases  in time according to a  \textit{cooling schedule}. 
Since SA solves the QUBO formulations of our problems, this algorithm was taken as  main comparison against Quantum Annealing.   


\section{Problem Complexity and Embedding}

\subsection{Resources Estimation}
\begin{figure}[t!]
	\centering
	\includegraphics[width=0.45\textwidth]{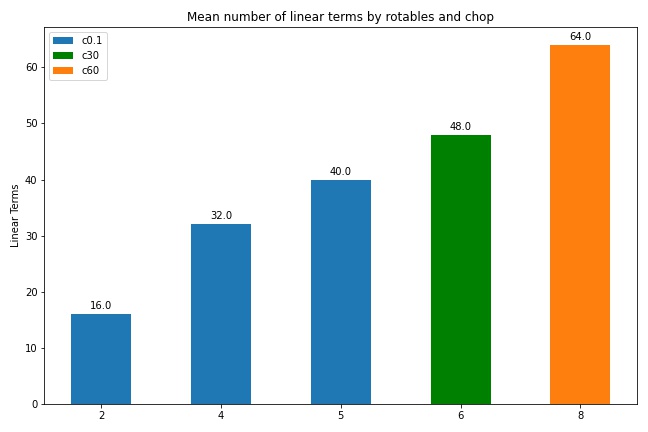}
	\includegraphics[width=0.45\textwidth]{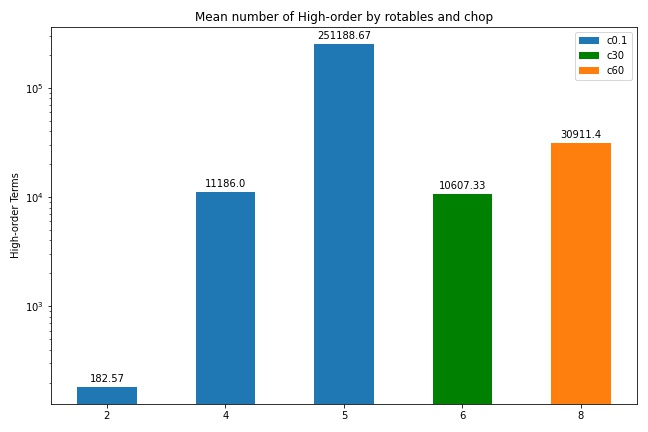}
	\caption{Mean number of linear and non-linear terms in the  HUBOs, obtained varying the number of torsionals and value of the approximation threshold .}
	\label{fig:capitolo6:avg_num_unapprox_hubo}
	
\end{figure}

The  metrics considered for evaluating the complexity of the problem are   the number of linear and high-order terms appearing in the HUBO  (plotted  in Figure \ref{fig:capitolo6:avg_num_unapprox_hubo}). The number of linear terms follows the relation  $Number\ of\ rotatables \times \ granularity \ of \ rotation$. Recall that  a granularity of 45 degrees was selected , i.e. $d=8$ in equation \ref{eq: discretization}.

The number of non-linear terms grow rapidly  going from 2 to 6 torsions. However, increasing problem size,   the  threshold approximation  allowed us to avoid the exponential growth that these numbers would otherwise follow.

Since the final objective is to be able to embed our problems into a QPU, a second threshold approximations acting on the QUBO has been applied. 
The choice of the amount of chop is done by a fine tuning of the threshold, which is lowered until the embedding cannot be performed anymore.

\begin{figure}[t!]
	\centering
	\includegraphics[width=0.45\textwidth]{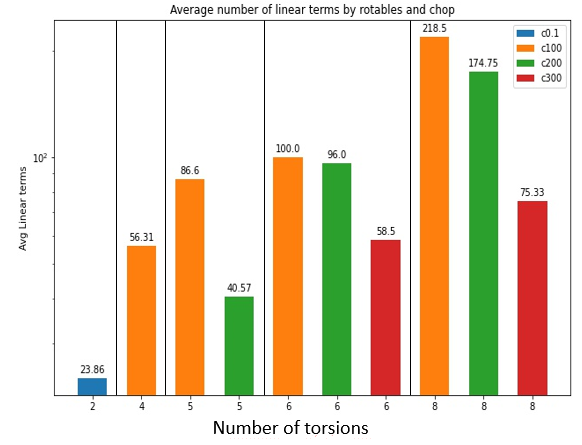}
	\includegraphics[width=0.45\textwidth]{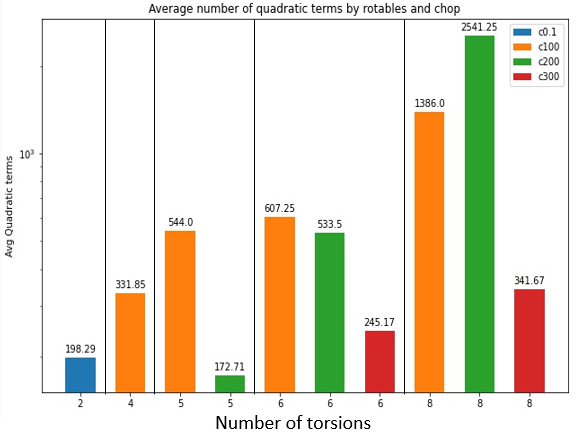}
	\caption{Average number of linear and quadratic terms in the  QUBOs, obtained varying the number of torsionals and value of the approximation threshold.}
	\label{fig:capitolo6:avg_num_approx_qubo}
\end{figure}

The plots in  Figure \ref{fig:capitolo6:avg_num_approx_qubo} show an increasing trend in the number of QUBO linear terms associated with  the same threshold, for an increasing number of rotatables. Linear terms also  decrease by increasing the threshold when the  number of torsionals is fixed.
Quadratic terms have a similar behaviour, except done for the class with  8 rotatable and a value for the approximating  threshold set at 200, where more complex problems  seem to peak.

\subsection{D-Wave Embedding}
%




In the embedding phase, a heuristic algorithm tries to find the optimal matching between problem resources and physical D-Wave hardware.
Several  embedding are possible, so the one that uses the least physical qubits out of few trials is employed.  It's implicit in this choice   that such embedding  will also have shortest chains on average, i.e. chains of physical qubits representing a single QUBO variable.

\begin{figure}[t!]
	\centering
	\includegraphics[width=0.45\textwidth]{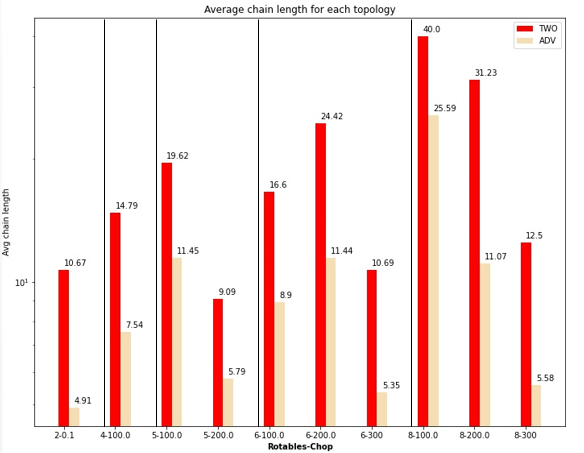}
	\includegraphics[width=0.45\textwidth]{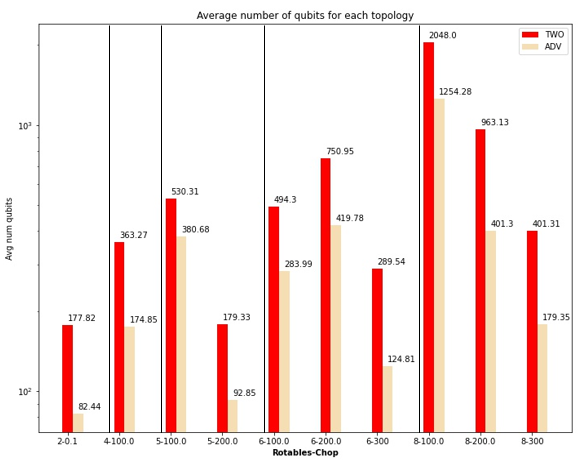}
	\caption{Plots depicting the variation of  average chain length and  average number of qubits in the embeddings by increasing  number of rotatables and value of the approximating threshold (also called chop)}
	\label{fig:capitolo6:avg_num_embeddings}
\end{figure}

The embeddings  shown in Figure \ref{fig:capitolo6:avg_num_embeddings} are highly representative of  the capabilities of each  QPU topology. 
On average, the chains obtained with a Chimera topology are 2.09 times longer than those found using the  Pegasus topology, which therefore  returns  52\% shorter chains.

Regarding the number of qubits, Pegasus is again the winner as it enables embedding with 1.39 up to 2.4 times less qubits with respect to Chimera.
On average, Advantage is 1.96 times more efficient than 2000Q, or  51\% less expensive in terms of qubits; this will be reflected in the quality of results.
There are only two results slightly diverging from our expectations. 
The  columns regarding the class $6-100$ are shorter than expected because, due to the complexity of the problem,   only  simpler instances were embedded. Hence the average is biased towards a lower value.

There other   important detail regards  the class  $8-100$ where  both  the average chain length and the average number of qubits are saturated to the maximum value for  the 2000Q Chimera topology. This happened because the device couldn't embed all the possible molecules for that class, therefore   radically increasing the average.

\section{Experimental Results}

A short remark should be made on the way these algorithms run.
Regarding the QA set-ups,  since the QUBO constant \emph{$A_{const}$}  modulating the strength of the hard constraint must be tuned, ten runs  for each value assigned to $A_{const}$ are performed. $A_{const}$ is evaluated as the maximum coefficient in the optimization contribution multiplied by an increasing factor. Each run is performing ten thousands forward anneals with an annealing schedule of one microsecond. 
Simulated annealing is performed on the same QUBOs,   with the possibility of choosing the number of epochs and the function  which is responsible for the temperature   decrease. In our case we used 500 annealing epochs and a geometrical decrease function.

Since  executions of  quantum  and simulated annealing are repeated many times in the attempt of finding the best solution, also the other classical techniques are executed until their run-time does not exceed a  time limit which was set from the beginning.
Random search divides the solution space in N parallel processes and as long as it does not run out of time, it executes and records the maximal value found. 
The Geodock Search is trying an incremental approach, which consists in performing the rotation on  an incremental number of bonds.


\subsection{Volume Gain in Time}

\begin{figure}[htpb]
	\includegraphics[width=.48\textwidth]{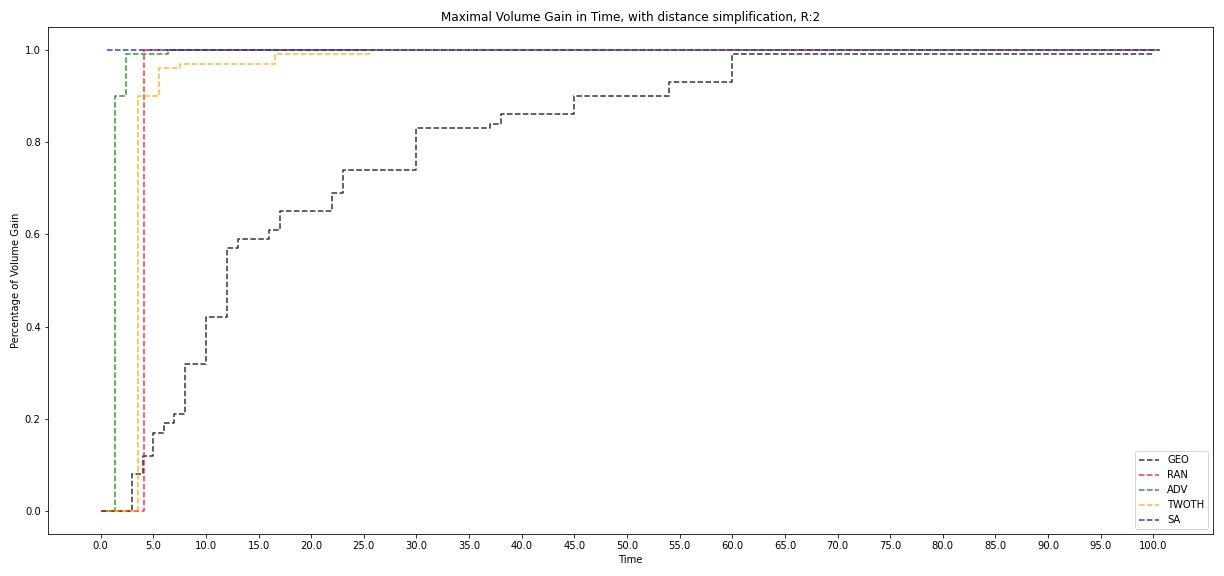}
	\includegraphics[width=.48\textwidth]{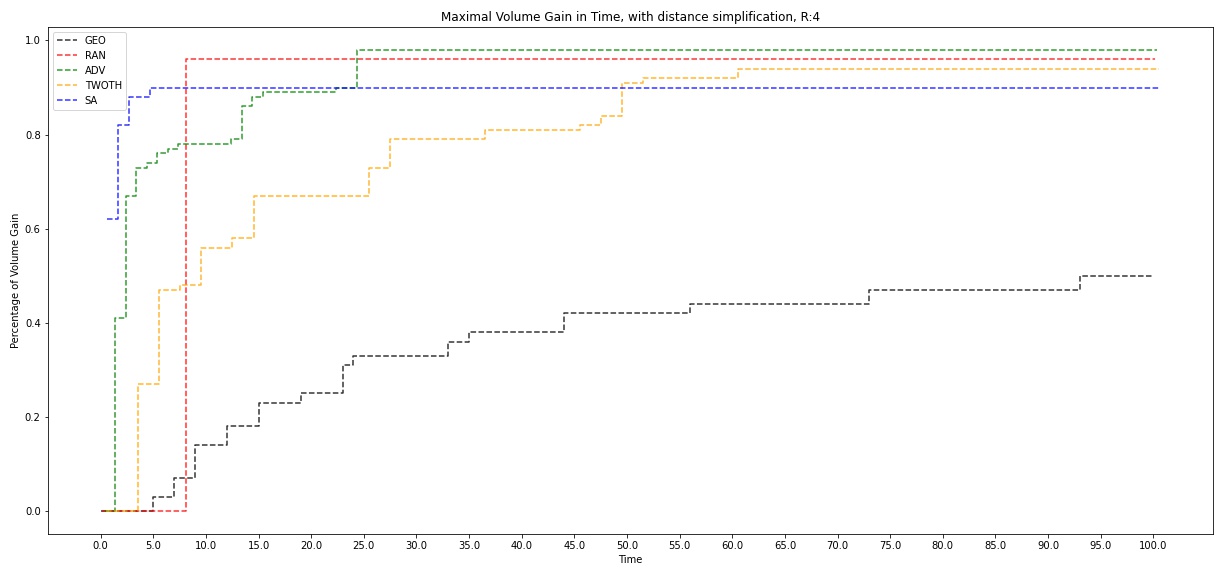}
	\includegraphics[width=.48\textwidth]{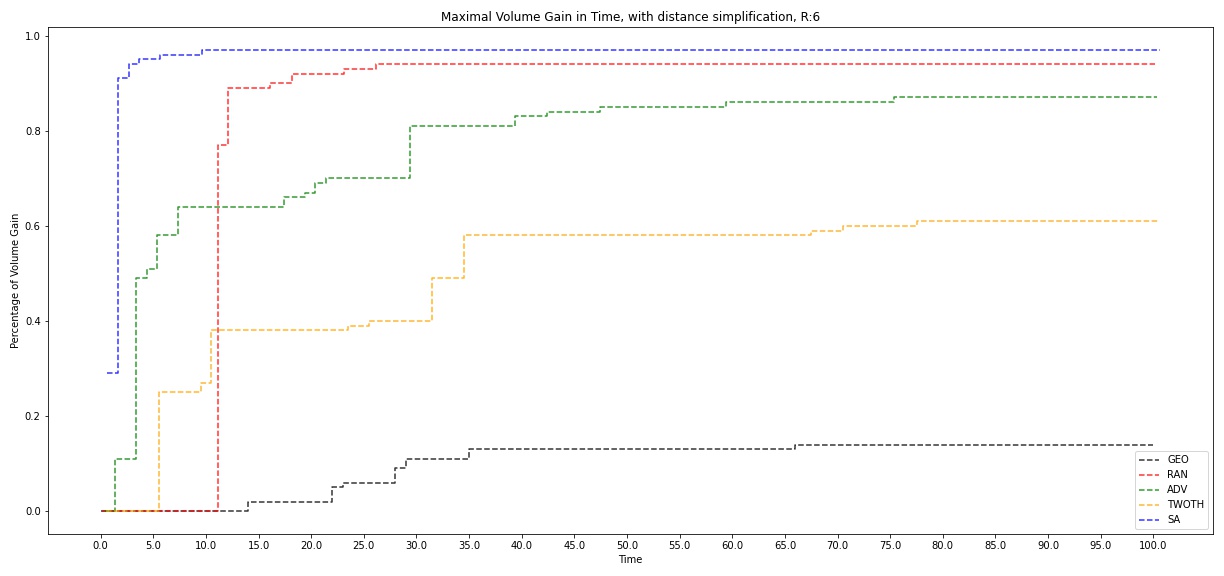}
	\includegraphics[width=.48\textwidth]{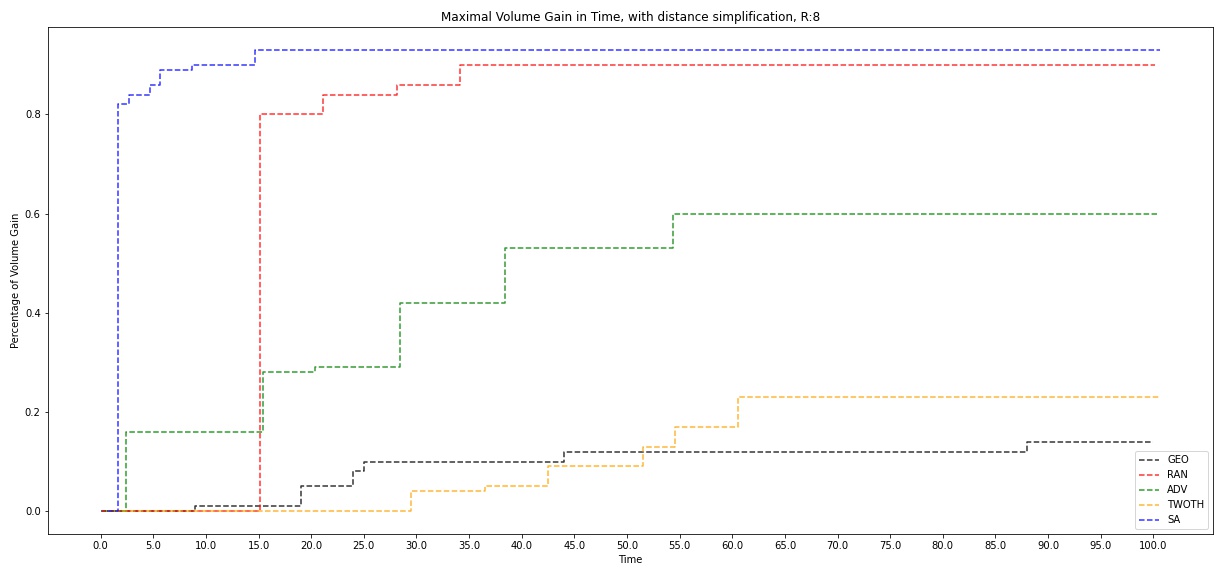}
	\caption{Comparison of average volume gained on a time window of 100 seconds for 2, 4, 6 and 8  rotatable bonds.}
	\label{fig:capitolo6:distance_true}
\end{figure}

Plots in the Figures \ref{fig:capitolo6:distance_true} show the average trends in volume gain obtained by each algorithm on a fixed time window, increasing  rotatable bonds.
On the Y-axis,  the  maximum volume reached by every algorithm at each second is displayed.  The plots show the algorithmic behaviour up to 100 seconds since  this is the interval where  greatest changes in volume appear.

First thing worth noticing is that  SA gave the best overall performance as it was able to provide good solutions falling inside the  $[90\%, 100\%)]$ volume gain range   with a run-time that never exceeded the 15 seconds.

It is also possible to observe that Advantage is the best among   D-Wave devices. For an increasing number of rotatable bonds,  volume achieved by 2000Q quickly dies off, while the results of Advantage remain competitive for a good number of torsionals.

When 2 torsionals are considered,  the MU problem looks like an easy task and  basically all the approaches are able to recover the optimal solution (only GeoDock gets close but fails to find optimal). Random search, SA and the Advantage QA are able to reach the  100\% volume gain within the first 5 seconds of runtime.

Increasing the problem size to 4 torsionals we can foresee a deviation of the trends: SA is again the fastest to reach a plateau while  the quantum annealers perform better than random search in the first 5 seconds. After that, random search presents a disruptive improvement which comes  with a delay due to  overheads in the memory access. 
A promising result is obtained after 25 seconds, where Advantage is able to find the optimal solution among the methods. This  indicates that QA could be a  competitive approach when  solving  medium-small sized problems.

It is possible to see another  reversal of the trends when the number of torsions rises to 6, in which SA and random search establish themselves as the most reliable solvers, followed by the Advantage quantum annealer. This behaviour is confirmed at 8 rotatable bond, where performances of 2000Q become very similar to that of  the GeoDock greedy algorithm.

\subsection{Comparison between SA and  QA}

Comparing classical algorithms with quantum annealers in terms of absolute time isn't fair, since the quantum devices have run-times that are comprising of  programming time,  readout time and  delay time, besides the time spent in actually performing the annealing.

Therefore  more suitable metrics for comparing annealing solvers is given by the  Time To Solution (TTS),
\begin{equation}\label{eq:TTS}
TTS = \frac{total \ execution \ time}{occurrence \ of \ the \ best \ solution}
\end{equation}
The total execution time can be calculated as the number of anneals multiplied by  the total  access time for  single anneal.   The occurrences of the best solution simply count  the number of times the best solution is found in the annealing.
The TTS metric can be interpreted as the inverse of the probability of finding the optimal configuration in a unit time interval. Moreover, the TTS is telling us how long we have to wait on average before the annealer outputs the best solution. For this reason, better solvers have  low values of TTS.

\begin{figure}
	\centering
	\includegraphics[width=0.8\linewidth]{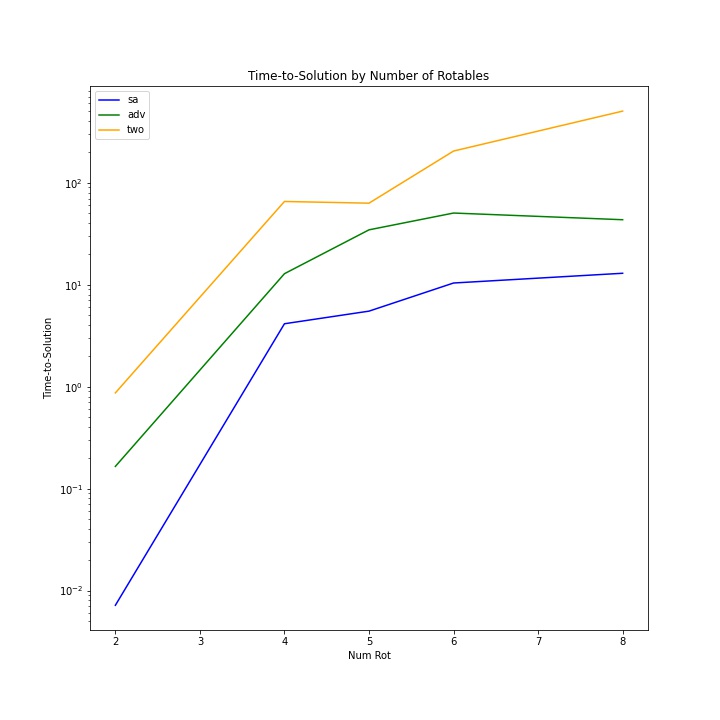}
	\caption{A comparison between Simulated Annealing, Advantage and 2000Q on TTSs, for increasing number of torsionals.}
	\label{fig:capitolo6:comparisons_tts}
\end{figure}

Plot in figure \ref{fig:capitolo6:comparisons_tts} show that the  TTS is very high for 2000Q, followed by Advantage, while SA has the the best TTS. Simulated annealing has a TTS several times smaller than Advantage, which means that its convergence to the optimum is much faster than in quantum annealing.

\begin{table}
	\centering
	\caption{Comparison on the TTS, for increasing number of torsionals.}
	\begin{tabular}{rrrr}
		\toprule
		torsionals &         sa &        adv &         two \\
		\midrule
		2 &   0.007171 &   0.165081 &    0.868552 \\
		4 &   4.138503 &  12.818871 &   65.482955 \\
		5 &   5.502274 &  34.508321 &   62.996198 \\
		6 &  10.384806 &  50.430784 &  204.457250 \\
		8 &  12.942243 &  43.363135 &  503.938826 \\
		\bottomrule
	\end{tabular}
	\label{tab:caption}
\end{table}%

\begin{figure}
	\centering
	\includegraphics[width=0.8\linewidth]{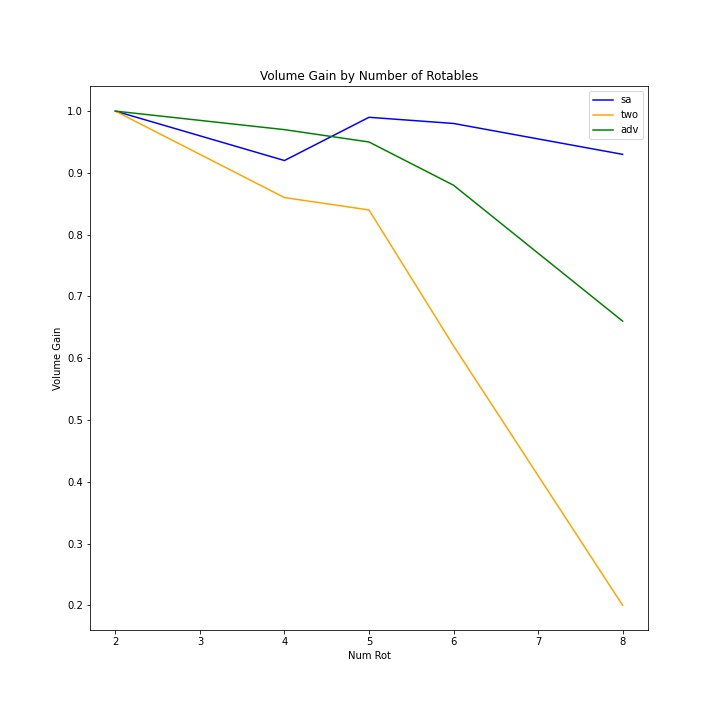}
	\caption{A comparison between Simulated Annealing, Advantage and 2000Q on volumes.}
	\label{fig:capitolo6:comparisons_vol}
\end{figure}

Regarding the volumes achieved, plots in Figure~\ref{fig:capitolo6:comparisons_vol} represents the degradation in the final volume gain as the problem complexity increases. The volumes  have a  common starting point since all the three algorithm actually reach the maximum steadily and completely for problems involving 2 rotatable bonds.
As seen in the previous section, Advantage works well up to 4 torsions,  SA has an inflection on 4 bonds but then recovers and finishes as the best in absolute terms; it has a volume which is 4.65 times better than the one found by 2000Q at 8 bonds, and 1.4 times better than  Advantage.

\begin{table}
	\centering
	\caption{Comparison on the Volumes, for increasing number of torsionals.}
	
	\begin{tabular}{rrrr}
		\toprule
		torsionals &    sa &   two &   adv \\
		\midrule
		2 &  1.00 &  1.00 &  1.00 \\
		4 &  0.92 &  0.86 &  0.97 \\
		5 &  0.99 &  0.84 &  0.95 \\
		6 &  0.98 &  0.62 &  0.88 \\
		8 &  0.93 &  0.20 &  0.66 \\
		\bottomrule
	\end{tabular}
	\label{tab:caption}
\end{table}%
The last concept worth mentioning is the measure of how much volume in percentage  can be gained per  unit of TTS. 
Practically, it is computed as the ratio between the normalized  volume gain and  TTS.

\begin{figure}[t!]
	\centering
	\includegraphics[width=0.4\textwidth]{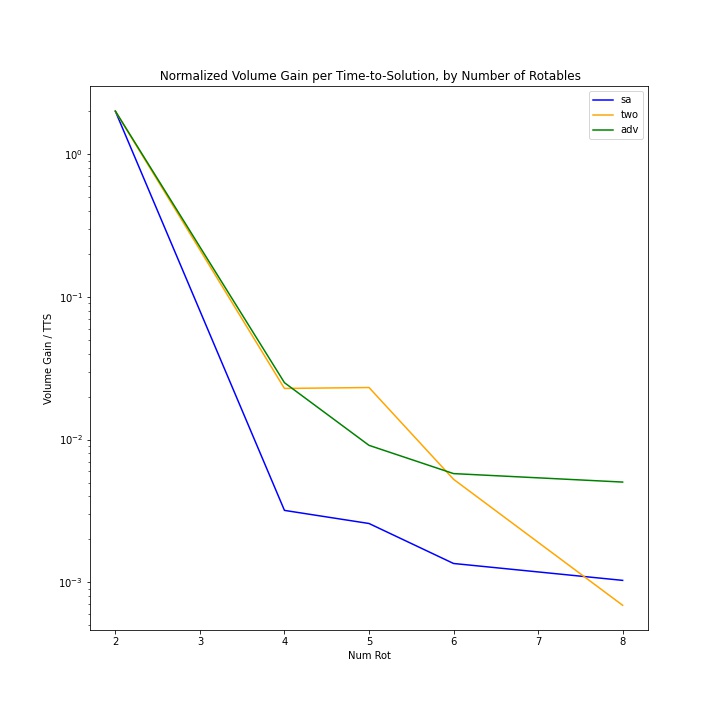}
	\caption{A comparison between Simulated Annealing, Advantage and 2000Q on the normalised decreasing quality of the volumes per time to solution.}
	\label{fig:capitolo6:comparisons_velocity_normal_velocity}
\end{figure}

A negative slope in the plot of figure \ref{fig:capitolo6:comparisons_velocity_normal_velocity} can be interpreted  as an increasing TTS required to achieve the same volume when scaling up to a more difficult problem.
A higher negative slope means  faster worsening of the algorithm with the increasing difficulty, while a smaller slope is a synonym of stability. Therefore, trends   running towards the bottom of the plot  belong to  algorithms which are defective.
The SA line falls down very fast from 2 to 4 torsionals, and then starts becoming stable. The same happens to Advantage which slows down this negative acceleration before. 2000Q never stops decreasing its quality, if not for a special case between 4 and 5 torsionals.

By checking the ratio of the slopes when increasing the number of torsionals, we can understand how much the algorithms are worsening adding  degrees of freedom.
On average the smallest growth in scaling of the slopes belongs to Advantage, with 1.36, whereas SA presents a higher 1.609, meanwhile 2000Q has an average of 6.01.

Although the results obtained with  Advantage may not be the winners in absolute terms, the fact that Advantage has   good performances  with respect to SA according to this metric  is a very promising.  This result  reflects how much QA is still immature  but also how far from saturation the margin of growth for QA really is.   Improvements in the next generation of QPU hardware could definitely affect both TTS and quality of sampling in a positive way.

\begin{table}[htp]
	\centering
	\caption{Average Normalized Volume Gain over TTS slopes  for the three annealing algorithms. }
	\begin{tabular}{rrr}
		\toprule
		sa &       two &       adv  \\
		\midrule
		1.609948  & 6.0189816 &  1.36206524  \\
		
		\bottomrule
	\end{tabular}
	
	\label{tab:caption}
\end{table}%


\begin{table}[t!]
	\centering
	\caption{Normalized Volume Gain over TTS for each algorithm, at different number of rotatable bonds.}
	\begin{tabular}{rrrr}
		\toprule
		sa &       two &       adv &  torsionals \\
		\midrule
		2.001031 &  2.000690 &  2.005038 &   2 \\
		0.003190 &  0.022822 &  0.025046 &   4 \\
		0.002582 &  0.023171 &  0.009112 &   5 \\
		0.001354 &  0.005269 &  0.005776 &   6 \\
		0.001031 &  0.000690 &  0.005038 &   8 \\
		\bottomrule
	\end{tabular}
	
	\label{tab:caption}
\end{table}%

\section{Conclusions}

In this work, we explored the possibility of  using a quantum annealer to support the drug discovery process within the in-silico virtual screening phase.
In particular, we studied a new method regarding the process of molecular unfolding, which is the first step in geometric molecular docking techniques.
This phase aims at finding the molecule's configuration that maximizes its volume, or equivalently, that maximizes the internal distances of the atoms that compose the molecule. 
We proposed our Quantum Molecular Unfolding model with the aim of executing it on D-Wave's latest hardware, Advantage and 2000Q.
The usage of the quantum annealers has the objective of understanding the capabilities of these quantum annealing devices and discovering if it is possible to improve the quality and the throughput of the ligand expansion in the state-of-art molecular docking methods.

The model is constructed starting from the identification of the rotatable bonds which are the parameters of the problem.  Once their discrete rotations are rewritten via  one-hot encoding thanks to the introduction of binary variables, the total sum of  internal atomic distances can be expressed as a HUBO. The HUBO is then transformed into a QUBO after   three steps are applied: (i) distance simplification, (ii) coarse-grained rotations  and (iii)  threshold approximation. 
These three approximations enabled us to reduce the complexity of the model with the aim of  embedding and running complex instances on the  QPUs.

We compared the performances of Advantage and 2000Q, with three other optimization methods, which are parallel random optimization,  simulated annealing, and a GeoDock greedy algorithm. The outcomes of the experiments show that Advantage have good performances on medium-small  problems while  classical techniques are  better than  quantum approaches when problem size increases.   Interestingly, the results obtained by the quantum annealer, both while running on Advantage and 2000Q, are better than the GeoDock greedy approach. 
In terms of the evolution of the quantum annealers, embedding our problems on Advantage, compared to 2000Q, was on average 51\% less expensive in terms of qubits and with chains 52\% shorter. Advantage significantly outperforms 2000Q in terms of time-to-solution (TTS) and volume gain when increasing the number of torsionals. Advantage also presents the best scaling of the normalized volume gain over time-to-solution when compared to both 2000Q and simulated annealing; this last result is one of the most promising as D-Wave devices are only bound to get better.

Future work can be the introduction of new dynamic thresholds approximation, which may lead to smaller embeddings but still of high quality, and  the introduction of   hybrid quantum approaches. A more difficult but also feasible and interesting task could be  extending  the quantum annealing approach to the whole docking process.
Finally, we are confident that this study will shed further light on the applications of quantum computing, thus enriching knowledge on topics that are becoming central in computational sciences.

\nocite{*} 
\bibliographystyle{unsrt}

\bibliography{sample.bbl}

\section*{Appendix}
\subsection*{Machines and libraries}

The platform used for running the classical algorithms is based on NUMA nodes, featuring 2 Intel(R) Xeon(R) CPU E5-2630 v3 CPUs (@2.40GHz), Virtualisation VT-x, caches L1d, L1i cache of 32K. L2 and L3 caches are, respectively, 256K and 20480K. RAM memory of 125 Gb and 114 Gb of SWAP memory. Operative system used is Ubuntu 18.04.5 LTS Bionic. The code was written in python 3.9.1, compiled with gcc 9.2.0.
The parallelization was actuated with the usage of mpi4py 3.0.3, compiled with MPICH 3.3.2.

Since the classical probabilistic algorithms and random search, could be run in embarrassingly parallel mode, mpi4py~\cite{Dalcin2011} was used. For each problem instance, the number of processes used for each run was 32. 

Other libraries of support are: Pandas 1.2.1, Numpy 1.19.5,
symbolic calculations were performed thanks to Sympy 1.7.1~\cite{Meurer2017}. \\

\end{document}